\documentclass[]{aa}
\usepackage{psfig}
\begin{document}

\thesaurus{08(08.14.2)}

\title{Early spectroscopic observations of Nova (V1494) Aquilae 1999 No.2
\thanks{Based on the data obtained at the David Dunlap Observatory,
University of Toronto}}

\author{L.L. Kiss\inst{1} \and J.R. Thomson\inst{2}}

\institute{Department of Experimental Physics and Astronomical Observatory,
University of Szeged,
Szeged, D\'om t\'er 9., H-6720 Hungary \and
David Dunlap Observatory, University of Toronto, Richmond Hill, Canada}

\titlerunning{V1494 Aql}
\authorrunning{Kiss \& Thomson}
\offprints{l.kiss@physx.u-szeged.hu}
\date{}

\maketitle
 
\begin{abstract}

Low- and medium resolution spectra of the fast nova, Nova (V1494)
Aql 1999 No.2 obtained approximately 6, 7, 19 and 28 days after the
maximum brightness
are presented and discussed. The spectrum covering the whole optical
range at day 6 shows the principal plus
diffuse-enhanced spectrum. The presence of strong
Fe II multiplets with P-Cyg profiles suggest that
V1494~Aql belongs
to the ``Fe II'' class defined by Williams (1992).
The medium-resolution profiles ($\lambda /\Delta \lambda
\approx 7000$) of the H$\gamma$ and H$\delta$ lines shows
well-defined sharp absorption features with the same radial
velocities,
while the H$\alpha$ split into
two distinct emission peaks in the last two spectra ($\Delta t$=19 and
28 days). The observed behaviour suggests an expanding
equatorial ring with possible small-scale
clumpiness in the nova shell.
The visual lightcurve is used to deduce M$_{\rm V}$ by the
maximum magnitude versus rate of decline relationship.
The resulting parameters are: $t_2=6.6\pm0.5$ days, $t_3$=16$\pm0.5$
days, M$_{\rm V}=-8.8\pm0.2$ mag. Adopting this value, a distance
d=3.6$\pm$0.3 kpc is determined.

\keywords{stars: novae -- stars: individual: V1494~Aql}
 
\end{abstract}

\section{Introduction}

Nova~Aql 1999 No. 2 (=V1494~Aql) was discovered visually by
A. Pereira on Dec. 1.875 UT, 1999 (Pereira et al. 1999)
at magnitude m$_{\rm vis}$=6.0. The spectroscopic confirmation
was given by subsequent low-resolution observations revealing
hydrogen Balmer series with P-Cyg profiles, Mg II 448.1 nm (or
He I 447.1 nm) and O I 777.4 nm lines, all with P-Cyg profiles.
The very early spectra showed the H$\alpha$
absorption component to be blueshifted by about 1020 km~s$^{-1}$ in
respect to the emission peak (Fujii et al. 1999). In contrast to
this latter measurement, Moro et al. (1999) reported the
absorption component of the H$\alpha$ P-Cyg profile at a
blueshift of $-$1850 km~s$^{-1}$. Additionally, Fe II multiplets
at 492, 502 and 517 nm were observed in emission with P-Cyg profile.
Further low-resolution spectroscopy was reported by W. Liller (Liller
et al. 1999), who took a CCD spectrogram with an objective
prism, that showed H$\alpha$ in emission at a level of 31 percent
above the local continuum.

Early photometric observations consist of mainly visual estimates
carried out by amateur astronomers published partly in a number of
IAU Circulars and partly in observing reports collected by the
VSNET group ({\tt http://www.kusastro.kyoto-u.ac.jp/vsnet}).
Thanks to the fast access to the newly obtained magnitudes,
the realtime brightness evolution could be followed. The star reached
a maximum brightness of 4.0v shortly after the discovery, which
was followed by a rapid decline (see below). Few photoelectric
data were published which gave a similar picture to that based
on visual estimates. V1494~Aql was also detected at 0.85 and 0.45 mm
using SCUBA on the JCMT (Pontefract et al. 1999)

V1494~Aql is the brightest nova in the northern hemisphere
after Nova (V1500) Cygni 1975 -- Nova (V1974) Cygni 1992 was fainter
in maximum by about 0.2 mag, Warner 1995 --, therefore,
it provides a good opportunity to carry out thorough studies of a
nova explosion by various
instruments. Unfortunately, its celestial position disables
continuous follow-up observations in early 2000. Therefore,
the spectroscopic
evolution, especially the formation of the nebular spectrum
is difficult to monitor. The main aim of this paper is to
present spectra of the nova taken about a week, 19 and 28
days after maximum and to discuss the meaning of the
observed spectra. In addition, we also estimate the absolute
magnitude and the distance of the nova
given by the characteristic light curve parameters ($t_2$ and $t_3$).

\section{The spectroscopic observations}

The spectroscopic observations were carried out with
the Cassegrain-spectrograph attached to the 1.88-m telescope
of the David Dunlap Observatory (Richmond Hill, Canada). The
spectra were obtained on four nights in December, 1999.
The detector was a Thomson 1024 x 1024 CCD chip (with a
6 e$^-$ readout noise). A low resolution broad band spectrum
was taken on the first night of observations, while
we took only medium resolution spectra to study
the line profiles in detail on the following three nights.
The gratings, observed
spectral regions and the resolution at the central wavelength
are summarized in Table\ 1.

\begin{figure}
\begin{center}
\leavevmode
\psfig{figure=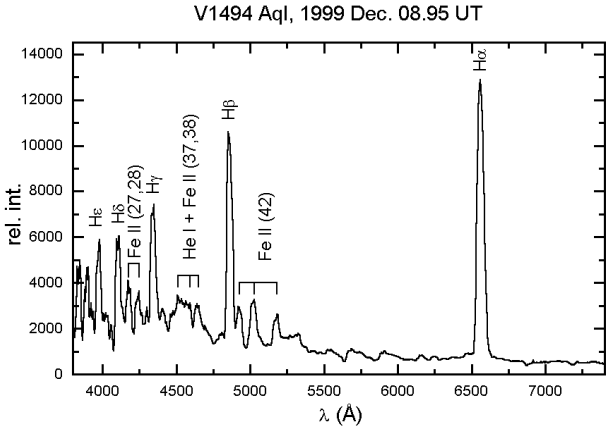,width=\linewidth}
\caption{The low-resolution spectrum of V1494~Aql}
\end{center}
\label{fig1}
\end{figure}

\begin{figure}
\begin{center}
\leavevmode
\psfig{figure=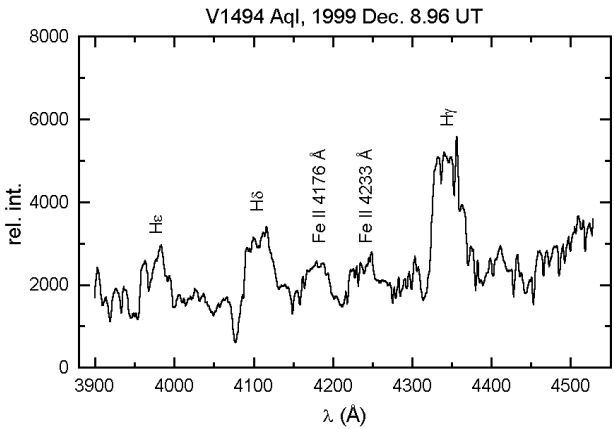,width=\linewidth}
\caption{The spectrum of V1494~Aql in the blue region}
\end{center}
\label{fig2}
\end{figure}

\begin{figure}
\begin{center}
\leavevmode
\psfig{figure=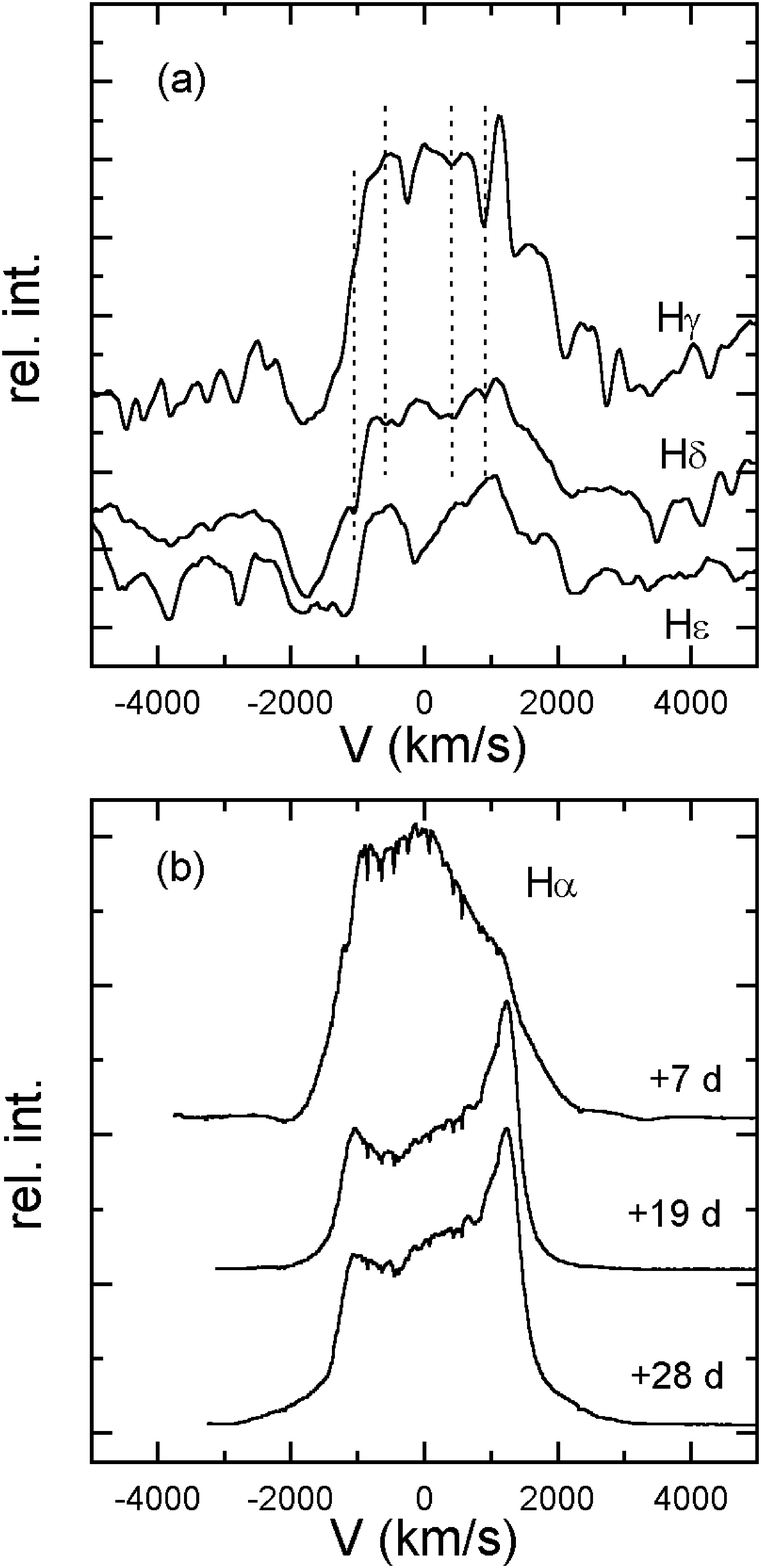,width=\linewidth}
\caption{Medium-resolution hydrogen profiles. (a): H$\gamma$,
H$\delta$ and H$\epsilon$, $\Delta t=6$ days; (b):
H$\alpha$, $\Delta t$=7, 19 and 28 days.
The vertical dotted lines in the top panel indicate the
subsystem of narrow absorption features at $\pm$400 and $\pm$1000
km~s$^{-1}$ (approximate values). The absorption components of
the P-Cyg profiles are blueshifted by $-1810\pm50$ km~s$^{-1}$ (H$\gamma$),
$-1790\pm60$ km~s$^{-1}$ (H$\delta$), $-1900\pm100$ km~s$^{-1}$ (H$\epsilon$)
and $-2000\pm100$ km~s$^{-1}$
(H$\alpha$). Note, that H$\alpha$ is affected by the
sharp atmospheric telluric lines. The splitted H$\alpha$ profiles
have a blueshifted peak at $-$1050 km~s$^{-1}$ and a redshifted
peak at +1200 km~s$^{-1}$}
\end{center}
\label{fig3}
\end{figure}

\begin{table}
\begin{center}
\caption{Journal of spectroscopic observations}
\begin{tabular} {llll}
\hline
Date (UT)  &  grating & range (\AA) & $\lambda / \Delta \lambda$\\
           &  (l/mm)  &             &                           \\
\hline
1999 Dec. 8.95 & 100  &  3750--7440 & 1400\\
1999 Dec. 8.97 & 600  &  3900--4500 & 7000\\
1999 Dec. 9.97 & 1800 &  6480--6680 & 11000\\
1999 Dec. 22.00 & 1800 & 6480--6680 & 11000\\
1999 Dec. 30.96 & 1800 & 6480--6680 & 11000\\
\hline
\end{tabular}
\end{center}
\end{table}

The spectra were reduced with standard IRAF tasks, including bias
removal, flat-fielding, aperture extraction (with the task {\it doslit})
and wavelength calibration. For the latter, two FeAr spectral
lamp exposures were used, which were obtained immediately
before and after every stellar exposures. The integration
times varied between 10 seconds and 10 minutes, depending
on the actual resolving power and wavelength range.

The obtained low-resolution spectrum is plotted in Fig.\ 1, where the most
characteristic features are also marked. Fig.\ 1
clearly illustrates the typical
principal plus diffuse-enhanced spectrum. The prominent
Balmer series and Fe II multiplets have P-Cyg profiles
that were already reported by Moro et al. (1999). Note the
Fe II and He I blend at 4500 \AA. The presence of
strong Fe II multiplets suggest that V1494~Aql belongs to
the ``Fe II'' class of classical novae defined by Williams (1992).
Adopting the interpretation of Williams (1992), this means that
the line formation happens in a post-outburst wind.
The higher-resolution view of the blue region is presented
in Fig.\ 2, where the hydrogen profiles show interesting
substructures in absorption. This
system of narrow absorption components of H$\gamma$ and
H$\delta$ lines can be seen in
Fig.\ 3a. The symmetric sharp absorption lines are shifted
approximately by $\pm$400 and $\pm$1000 km~s$^{-1}$ relative
to the central emission near the rest wavelength. H$\epsilon$
is shown for comparison, because of the similar complexity.

The evolution of the H$\alpha$ profile was followed through
4 weeks after the maximum. The earlier non-Gaussian
emission peak changed to a saddle-shaped profile
with two maxima at $-$1050 km~s$^{-1}$ and +1200 km~s$^{-1}$.
(Fig.\ 3b). Unfortunately, the possible similar narrow absorption
lines are strongly affected by the atmospheric telluric lines.

\section{The light curve}

To estimate the light curve parameters, we used
visual magnitude estimates by amateur astronomers that were
made available by the VSNET group (this data
also includes the observations published in the IAU Circulars).
Between the
discovery on Dec. 1, 1999 and Jan. 5, 2000, almost
420 individual points were collected. As the typical
uncertainty of a visual estimate is about $\pm$0.3 mag,
we have determined a mean light curve by taking 0.1-day bins
and calculating the mean value from the individual
points. This led to an averaged light curve
containing 117 points, which was further noise-filtered
by a simple Gaussian smoothing with 0.08 days FWHM. This
approach can only be used when the original light curve
is dense enough and therefore, the observational scatter
can be properly averaged (see Kiss et al. 1999 illustrating
this method).
The resulting visual light curve is plotted in Fig.\ 4, where
the maximum, $t_2$ and $t_3$ are also indicated.
The derived parameters are: $t_0$=2451515.9$\pm$0.1 days (1999
Dec 3.4 UT),
$t_2$=6.6$\pm$0.5 days, $t_3$=16$\pm$0.5 days. Consequently,
V1494~Aql is a fast nova and
the spectroscopic observations were done 6, 7, 19 and 28 days
after the maximum. We note,
that the statistical relationship between $t_2$ and $t_3$
($t_3 \approx 2.75 t_2^{0.88}$, Warner 1995) agrees well
with the observed values, as the determined $t_2$ would
imply a $t_3$ of 14.5 days.

The most striking post-maximum feature is the additional cyclic
brightness change following the fast decline. The characteristic
period of this secondary variation is about 7 days with
an approximate semi-amplitude of 0.3--0.4 mag. It occured
10 days after the maximum light. Fortunately, the nova was at 7.0 mag
(maximum plus 3 mag) around the mean level of the cyclic change,
therefore, $t_3$ is only weakly affected. The occurence of
the quasi-periodic brightness oscillations
indicated the start of the transition phase, when the
fast novae often show variations with quasi-periods ranging
from 5 days (GK~Per) to 25 days (DK~Lac, Warner 1995), generally
with a time-scale being close to $t_2$. The underlying
physical mechanism could be related either to accretion disk
phenomena (Leibowitz 1993) or to the contracting photosphere
of the white dwarf (Bianchini et al. 1992).

\begin{figure}
\begin{center}
\leavevmode
\psfig{figure=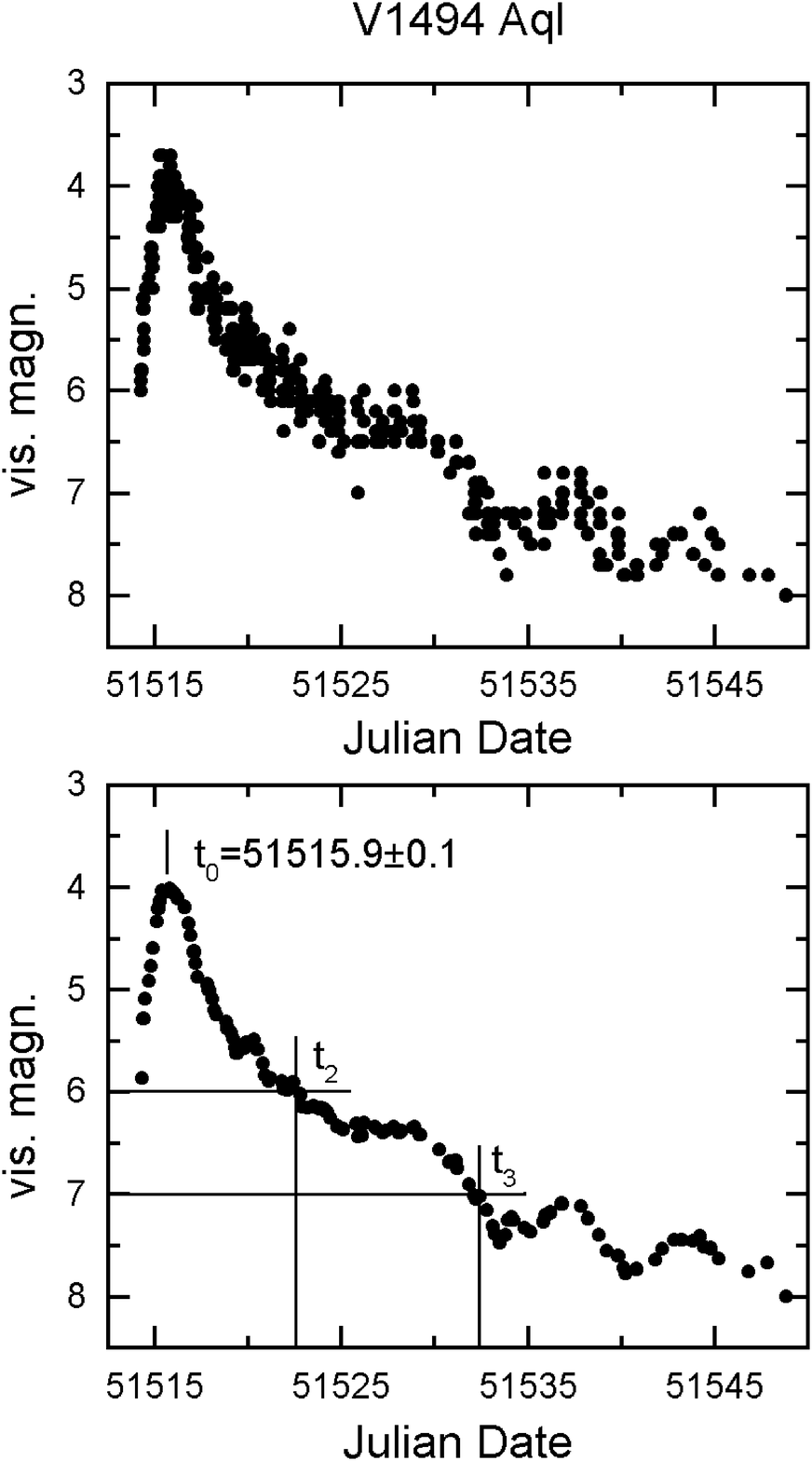,width=\linewidth}
\caption{The visual light curve of V1494~Aql}
\end{center}
\label{fig4}
\end{figure}

Three maximum magnitude versus rate of decline (MMRD) relations
were used to calculate visual absolute magnitude (Della Valle
\& Livio 1995, Capaccioli et al. 1989 and Schmidt 1957).
They gave $-$8.84 mag, $-$8.87 mag and $-$8.72 mag
for M$_{\rm V}$ and M$_{\rm vis}$, respectively. Another method, involving
the absolute magnitude 15 days after maximum (Capaccioli
et al. 1989), resulted in M$_{\rm V}$(max)=$-8.5\pm0.15$ mag (formal
error), however, this value has higher uncertainty due to
the brightness oscillation in the transition phase.
All of these values suggest an approximate visual absolute
magnitude of $-8.8\pm0.2$ mag being typical among the fast
novae. The distance can be determined from the apparent
maximum brightness (m$_{\rm vis}$=4.0 mag). Neglecting the interstellar
reddening, the distance is d=3.6$\pm$0.3 kpc. The absence
of strong Na I D absorption may suggest that the reddening
is low indeed.
The mass of the white dwarf
in V1494~Aql was estimated using Eq. (6) of Livio (1992)
assuming $(B-V)_0^{max} \approx 0$ (i.e. M$_{\rm B}^{max} \approx
M_{\rm V}^{max}$)
(Warner 1995).
It resulted in M$_{\rm WD}$=1.1 M$_\odot$, however, this
value should be only considered as a rough estimate.
Finally, the calculated distance and the apparent
brightness of the likely progenitor (Pereira et al. 1999)
give the pre-outburst visual absolute magnitude of the system.
There is a star in the USNO-A2.0 catalogue very close to
the position of the nova with red mag 15.6 and blue mag 17.4.
These values correspond to an approximate visual apparent magnitude
16.5, that results in M$_{\rm vis}$=3.7 mag for the
progenitor, which is normal for quiescent novae (Ringwald
et al. 1996). This result excludes the possibility of
a giant secondary component, like in GK~Per or RS~Oph.

\section{Conclusions}

The overall appearence of the low-resolution
spectrum covering the whole optical range strongly
resembles Nova LMC 1988 No.1 (see Fig.\ 1 in
Williams 1992) being a nova with characteristic ``Fe II''
type spectrum. Also, it is very similar to
the spectrum of the fast ($t_2$=5 days) nova,
Nova LMC 1988 No. 2 (Sekiguchi et al. 1989, Williams et al. 1991).
Sekiguchi et al. (1989) found two distinct sharp absorption
systems in the Balmer lines resembling our medium-resolution
observations of the H$\gamma$ and H$\delta$ profiles.
The presence of such complex absorption systems
is a
common phenomenon in novae (Sekiguchi et al. 1989),
which may usually be traced even to H$\epsilon$--H8 in the Balmer series.
Unfortunately, we have obtained only one spectrum in that
spectral region, therefore, we have no information
on the time evolution and no firm conclusion can be
drawn about the underlying physical reasons.
One would speculatively expect such symmetric ($\pm$400
and $\pm$1000 km~s$^{-1}$) and narrow absorption lines
in an axisymmetric wind during the common envelope (CE)
phase in the outburst, when the CE phase produces
a density contrast between the equatorial and polar
directions (Livio et al. 1990). This is supported
by the observed H$\alpha$ profiles, which
showed first an emission profile with one main component (simultaneously
with the H$\gamma$ and H$\delta$ observations) and
then the split into two main peaks after 19 days.

The saddle-shaped H$\alpha$ profile suggest
the non-spherical symmetry of the nova shell. Most
recently, Gill \& O'Brien (1999) presented an
ensemble of calculated emission-line profiles from model
nova shells with various symmetries. The observed shape
of the H$\alpha$ line corresponds to an equatorial ring
seen most probably nearly edge-on. The difference
of the blue and red peaks in the H$\alpha$ can
be associated with possible small-scale
clumpiness in the shell, which produces an
increase in brightness in the receding half of the shell
(see Gill \& O'Brien 1999 concerning the observed
asymmetries in the emission-line profiles of V705~Cas).

Adopting the calculated distance of 3.6 kpc and
and expansion velocity of the nova shell of 2000 km~s$^{-1}$
given by the radial velocity of the absorption component
of the P-Cyg profiles, Eq. (5.10) in Warner (1995)
gives 0\farcs11 for the angular radius of the shell
after 1 year. This is within range of the HST. Therefore,
subsequent spectroscopy and high-resolution imaging
may confirm the existence of the hypothetic
non-spherical structure. Evidence suggests this to
be an equatorial ring.

\begin{acknowledgements}
This research was supported by Hungarian OTKA Grants \#F022249,
\#T022259 and Szeged Observatory Foundation.
Fruitful discussions with J. Vink\'o are gratefully
acknowledged. The referee (Dr. M. Orio) has greatly
improved the paper with her helpful comments and
suggestions. The NASA ADS Abstract
Service was used to access data and references.
\end{acknowledgements}

\end{document}